\documentclass[conference]{IEEEtran}
\IEEEoverridecommandlockouts

\usepackage{cite}
\usepackage{amsmath,amssymb,amsfonts}
\usepackage{graphicx}
\usepackage{textcomp}
\def\BibTeX{{\rm B\kern-.05em{\sc i\kern-.025em b}\kern-.08em
    T\kern-.1667em\lower.7ex\hbox{E}\kern-.125emX}}

\usepackage[table,dvipsnames]{xcolor}
\usepackage[T1]{fontenc}
\usepackage{url}
\usepackage{booktabs}
\usepackage[acronym,nomain,nonumberlist]{glossaries}
\usepackage{sc26repro} 
\newcommand{\code}[1]{\texttt{#1}}

\newacronym{ai}{AI}{artificial intelligence}
\newacronym{aiml}{AI/ML}{artificial intelligence and machine learning}
\newacronym{ci}{CI}{continuous integration}
\newacronym{bsp}{BSP}{bulk synchronous parallel}
\newacronym{mpi}{MPI}{Message Passing Interface}
\newacronym{hpc}{HPC}{high performance computing}
\newacronym{aws}{AWS}{Amazon Web Services}
\newacronym{gke}{GKE}{Google Kubernetes Engine}
\newacronym{aks}{AKS}{Azure Kubernetes Service}
\newacronym{eks}{EKS}{Elastic Kubernetes Service}
\newacronym{ml}{ML}{machine learning}
\newacronym{rdma}{RDMA}{Remote Direct Memory Access}
\newacronym{os}{OS}{operating systems}
\newacronym{vm}{VM}{virtual machine}
\newacronym{llm}{LLM}{Large Language Models}
\newacronym{fom}{FOM}{figure of merit}
\newacronym{efa}{EFA}{Elastic Fabric Adapter}
\newacronym{ec2}{EC2}{Elastic Compute Cloud}
\newacronym{ucx}{UCX}{Unified Communication X}
\newacronym{cni}{CNI}{container networking interface}
\newacronym{ebpf}{eBPF}{extended Berkeley Packet Filter}
\newacronym{gvnic}{gVNIC}{Google Virtual NIC}
\newacronym{uri}{URI}{unique resource identifier}
\newacronym{api}{API}{application programming interface}
\newacronym{dag}{DAG}{directed acylcic graph}
\newacronym{crd}{CRD}{custom resource definition}
\newacronym{mcp}{MCP}{Model Context Protocol}
\newacronym{hci}{HCI}{Human-Computer Interaction}
\newacronym{pid}{PID}{Process IDentifier}

\begin{document}
\bstctlcite{IEEEtran:BSTcontrol}

\title{Descriptive Dispatch of Computational Work}

%\author{\IEEEauthorblockN{Anonymous Author(s)}
%\IEEEauthorblockA{\textit{Affiliation omitted for double-blind review}}}

 \author{\IEEEauthorblockN{Vanessa Sochat}
\IEEEauthorblockA{\textit{Lawrence Livermore National Laboratory} \\
Livermore, California, USA \\
sochat1@llnl.gov \\
ORCID 0000-0002-4387-3819}
\and
\IEEEauthorblockN{Daniel Milroy}
\IEEEauthorblockA{\textit{Lawrence Livermore National Laboratory} \\
Livermore, California, USA \\
milroy1@llnl.gov \\
 ORCID 0000-0001-6500-3227} }

\maketitle

\begin{abstract}
Agents powered by AI/ML are becoming ingrained in orchestration. Dispatch of work is the task of receiving a request, transforming it for a workload manager, and successfully submitting it. Running scientific workflows across multi-cluster environments introduces substantial challenges of dynamic job transformation, dispatch, and submission to heterogeneous clusters. These tasks are well-suited to agents, which can receive textual instructions for work, prepare job specifications, and dispatch. In this work, we assess the reliability of a dispatch agent across 432 runs, testing all possible combinations of five feature dimensions across four prompt styles. The agent is highly reliable (97.9\% success). We test a full orchestration to submit, queue, match, score, select, transform, and dispatch in a multi-cluster experiment. We find that descriptive metadata increases successful execution from 48\% to 87\% of 220 submitted jobs, eliminating architecture mismatch, and improving performance for five of ten measurable applications by up to 3.3x.
\end{abstract}


\section{Introduction}

Modern scientific workflows have moved beyond monolithic batch simulations toward integrated workflows that combine high-fidelity solvers with \gls{ai} and \gls{ml} models. The resources that are offered by any single cluster resource cannot always satisfy the needs of such a workflow, requiring portability to run across clusters or centers. This task is historically not easily possible, or needs manual work to transform jobs and rebuild applications for entirely different environments. 

% From other draft (last sentence above)
% Enabling dynamic portability has remained highly challenging, and typically requires human intervention to transform jobs and rebuild applications for different environments. 

% We might imagine a \gls{hpc} center of the future that changes this paradigm. By way of using \gls{llm}, a cluster or center resource might receive and intelligently respond to requests, handling tasks to transform the request for work, submit, monitor, and return updates to the calling user. In this possible future, a user no longer submits a job as a structured script with directives for a specific workload manager. It is provided as a textual request for resource quantity and type, and a preference for priority in time to completion, cost, and performance. This kind of initiative and monetary backing is also represented in national initiatives such as the Future Generation HPC Center (FG-HPCC) \cite{llnl_fghpcc_rfi_2025} and the Genesis Mission \cite{Gil2025Genesis}.

A \gls{hpc} center of the future will rely more heavily on \gls{ai} for automated application orchestration. Using \gls{llm}, a cluster or center resource might receive and intelligently respond to requests, handling tasks to transform a work request as well as submit, monitor, and return updates to the calling user. In this potential future, a user no longer submits a job as a structured script with directives for a specific workload manager. It is provided as a textual request for resource quantity and type, and a preference for priority in time to completion, cost, and performance. This kind of initiative and monetary backing is also represented in national initiatives such as the Future Generation HPC Center (FG-HPCC)~\cite{llnl_fghpcc_rfi_2025} and the Genesis Mission~\cite{Gil2025Genesis}.

Work in converged computing \cite{sochat2024converged} provides motivation for modeling how traditional \gls{hpc} technologies can be integrated with and extended to cloud-like environments \cite{sochat2025usability}. The movement serves as a means to transition from traditional techniques to new methods that improve automation, modularity, and discoverability. An \gls{hpc} center of the future requires automation, and automation demands that traditional workload managers and center resources be accessible programmatically, and specifically to agents that interact with \gls{llm}s. Anthropic's \gls{mcp} has surfaced as an open standard for a bidirectional client-server over JSON-RPC 2.0 that allows for agents to intelligently discover and use tools \cite{mcp}. Custom MCP servers have already shown promise in specialized tasks, including orchestrating quantum computing circuits on PBS schedulers \cite{Shiraishi2026-ca-quantum-execution} and extending to software development and the financial industry \cite{Stein2026-xo}.

If workloads are to be run across centers, agents can make scheduling choices based on resource availability and policy. Historically, multi-cluster scheduling has used specialized abstractions designed to bridge the gap between locations. In the traditional \gls{hpc} sector, \textit{Federated Slurm} has served as a mechanism for multi-cluster workload distribution, allowing a single submission point to route jobs across a global federation of clusters \cite{Bartkiewicz2016-mm, federated-slurm}. In cloud and Kubernetes, projects such as \textit{Volcano} have introduced high-performance batch scheduling capabilities, including gang scheduling and fair-share to Kubernetes~\cite{volcano2022}. To handle resource quotas and job queueing across clusters, \textit{Kueue} has emerged as a standard for cloud-native job management, providing a mechanism to queue jobs until resources become available \cite{kueue2024}. Furthermore, frameworks like \textit{Armada} and the \textit{Multi-Cluster App Dispatcher (MCAD)} offer higher levels of abstraction, enabling dispatching large-scale workloads across thousands of nodes spanning multiple availability zones and cloud providers \cite{armada2023, mcad2022}. The drawback of these approaches is that they often prioritize the first available or ready resource, which is not necessarily the best fit for a particular workload. 
% However, it is unclear if a single cluster resource and an agent secretary representing it can be reliable to dispatch.

Multi-cluster scheduling tools are extended to enable distributed workflows in cloud environments by other approaches such as round robin, lowest cost, capacity requests, and affinity-based ~\cite{wang2025efficient,bahreini2024caspian,el2026survey}. There is often a tradeoff between speed and optimization in that it is more costly to calculate a best fit. For this reason, the approach of first ready across clusters that could provide resource capacity is a balanced choice. However, first ready does not indicate faster completion. The completion time may be farther in the future than if it had run on more suitable resources or the cost could be greater. Furthermore, there are inefficiencies introduced by submitting jobs and subsequently canceling them when a best fit is determined. 
% compute resources are required to dispatch and cancel jobs, potentially loading all participating clusters. % An improvement to this approach would not blindly submit to all clusters, but rather select intelligently. 

While both resource provisioning and workflow management account for application-specific needs, the former falls under the domain of administration, hardware, and dispatch, while the latter focuses on application parameter selection, step coordination, and decision making. An \gls{hpc} center as a whole cannot be expected to have a single workload manager or software suite. It could contain a dynamic Kubernetes cluster running on cloud resources, a set of virtual machines with Slurm, a single server providing access to an edge device, or a traditional on-premises \gls{hpc} cluster. A work dispatch strategy cannot assume a specific manager, software distribution strategy, or network. The center resources must be probed, discovered, and made accessible not just to a local resource manager, but potentially to a fleet of computational resources that serves a broad set of users, use cases, and workload types.

% Not sure how to structure this - originally had ---
The full process to dispatch work includes negotiation, selection, and dispatch. Negotiation is an assessment for a resource to satisfy a request. Selection is choosing from contenders that can satisfy the request. Dispatch is the final step to deploy the work to the cluster. We have previously studied negotiation and selection \cite{sochat2026agenticscience} and in this work, we aim to run an experimental suite to assess the accuracy and reliability of agents for dispatch. Using a 5-node Kubernetes cluster, we evaluate the capability of a secretary agent \cite{resource_secretary_2026} to transform a high-level intent into a validated job specification for \emph{LAMMPS}. We make the following contributions:

\begin{itemize}
\item Secretary Agent software with dispatch capability
\item Experimental protocol to validate dispatch correctness
\item Association of prompt variants with success outcomes
\item Artifact Secretary to generate container manifests
\item Flux Secretary for application execution and monitoring
\item Fluxq software for end-to-end descriptive scheduling
\end{itemize}
In this work, we present a novel architecture and supporting software to enable the probe, discovery, proposal generation, and selection of workloads.

In Section \ref{sec:methods} we describe our controlled method for assessing agent reliability and accuracy of dispatch of work. We describe the extension to our resource secretary software to provision a dispatch agent. In Section \ref{sec:results} we analyze results, and conclude with discussion in Section \ref{sec:discussion}.

\section{Methods}
\label{sec:methods}

\subsection{Dispatch Experiments} 
\label{sec:dispatch-experiment-1}

We are interested in the extent to which an agent can successfully dispatch work to a cluster, which includes job translation, job submission, and success evaluation. Job translation is the conversion of a prompt or job specification from another manager to a target cluster, where success evaluation means ensuring the submission was successful and output was generated as expected. The agent needs to view sufficient log content to verify that the job started without error to return a successful status to the calling hub. The first challenge is loss of information in the initial translation, either due to options being different, being unsupported, or having no analogous transformation. The second challenge is the agent's ability to submit the job, monitor it, and return a response to the calling user.

\smallskip
\noindent{\bf Computational Resources} 
\label{sec:method-resources}
We chose to run our experiments in Kubernetes due to the declarative orchestration environment for state management and the ability to deploy heterogeneous clusters. Within the Kubernetes environment, we chose to deploy the Flux Framework workload manager \cite{Ahn2014-ku} to run \gls{hpc} applications. We ran experiments using a five node \texttt{hpc7g.16xlarge} Kubernetes cluster on \gls{aws} \gls{eks}. Using the Flux Operator \cite{Sochat2024-the-flux-operator} we deployed a LAMMPS container built with libfabric for \gls{efa} support. Each \texttt{hpc7g.16xlarge} node has 64 cores and 128 GiB of RAM.

\smallskip
\noindent{\bf Resource Secretary Software} 
\label{sec:resource-secretary-software} 
We use the resource secretary software \cite{resource_secretary_2026} to handle dispatch. The resource secretary implements over 50 real and simulated providers that range from workload managers to storage and network, and allows for a \gls{mcp} server to initialize, automatically discover available resources, and provide tools to an agent to further investigate the environment. The software supports ten workload managers, and we use Flux Framework for our experiments. For work dispatch, the Flux Framework workload provider is discovered, exposing tools for the agent to inspect the queue, manage jobs, and explore resources. The provider interface and \gls{mcp} server design has been described previously \cite{sochat2026agenticscience}. While the server can serve traditional \gls{mcp} functions for a scientific application, the provider tools are function handles called directly by the resource secretary, requiring no additional \gls{api} or server interaction.

\smallskip
\noindent{\bf Prompt Generation} 
\label{sec:prompt-generation}
The motivation for our investigation into prompt generation is to advise a human user how to best write prompts that will result in successful job execution. We aimed to improve upon our prompt generation strategy described in \cite{sochat2026agenticscience}, which used a specificity index to populate a template. Instead, we create an application-and provider- specific combinatorial prompt generator that creates combinations of the workload manager, presence of the \emph{nocite} flag, resources, application configuration, and affinity. The \emph{nocite} flag has no functional impact to LAMMPS. We use it as a proxy for asking an agent to use an arbitrary flag with different prompting styles. It serves as an additional check that the agent is following instructions and generating the intended command. We define four categories of prompts, and want to understand how successful each strategy is to run a job.  For each factor, we are able to define several variants for the underlying intent, including \emph{exact}, \emph{descriptive}, \emph{verbatim}, and \emph{discovery}.  An example for the Flux workload manager is the following:

\begin{itemize}
    \item \textbf{Exact:} Phrasing that mirrors client syntax (\texttt{flux run})
    \item \textbf{Descriptive:} Natural language descriptions (\emph{using the Flux workload manager}).
    \item \textbf{Verbatim:} Natural language descriptions (\emph{using flux run}).
    \item \textbf{Discovery:} Intentional ambiguity that requires the agent to explore the environment (\emph{run LAMMPS}).
\end{itemize}

We call this a \emph{provider vocabulary}, and the variants can be thought of as prompting strategies. We will want to understand the prompting strategies that are best suited for dispatch. We generate a matrix across 5 features and 4 variants for a total of 432 prompts. This granularity  will allow us to calculate an odds ratio to determine which combination of prompting strategy and variant leads to the best outcome, which is the highest success rate measure of the agent's ability to dispatch work. Each prompt is a template structured as follows: $\text{Prompt} = [\text{Manager}] + [\text{Resources}] + [\text{Application}] + [\text{Modifiers}]$.
% \begin{equation}
% \label{equation:features}
% \text{Prompt} = [\text{Manager}] + [\text{Resources}] + [\text{Application}] + [\text{Modifiers}]
% \end{equation}

We generate a matrix of prompts that represents all possible combinations of variants, and each prompt can be associated with a single command that represents a ground truth or correct submission. The simulation used a LAMMPS ReaxFF potential with a varied spatial domain. Specifically, we chose a problem size of $x=32$, $y=32$, and $z=16$. across 5 nodes for a total of 320 tasks.

\begin{small}
\begin{verbatim}
    flux submit -N 5 -n 320 lmp -v x 32 \
                -v y 32 -v z 16 -i in.reaxff
\end{verbatim}
\end{small}

In our resource secretary software, the specific modifiers for the workload manager and application are generated by the provider interfaces for each of Flux and LAMMPS. More specifically, a provider can be asked to generate a template for a specific feature (e.g., \emph{affinity}) and prompt style (e.g., \emph{discovery}) and return a textual string for the agent that goes into the final assembled prompt. We did each experiment twice. 

\smallskip
\noindent{\bf Agentic Job Dispatch} 
\label{sec:agentic-dispatch}
Job dispatch is requested when a specific prompt is sent to the agent. The agent is required to transform the request to be suited for a specific workload manager. For our experiments we use Flux Framework.  The agent is instructed how to make calls to the resource provider and application query interfaces.  A cluster running a different workload manager would require the agent to transform the equivalent request to a different resource provider interface. We allow a maximum number of 10 attempts to complete the entire orchestration, and require the agent to make at least two observations of the system. We chose two observations to reflect the minimum need for an agent to submit the job and check on a status. Not requiring observations can lead to hallucination. If the agent attempts to return a result without meeting this threshold, we remind the agent to verify its actions through tool-based observations. The agent is told that it is required to use a validation tool to check LAMMPS parameters before submission. In the case of an incorrect problem size or incorrectly formatted arguments, the agent is instructed to fix the command. We added this check to ensure that when we asked for an exact problem size, the agent would not decide to change it or format it incorrectly. In practice this check reduces erroneous runs and ensures the problem is run as intended. The agent is required to generate the job specification, submit it to the workload manager, check the status and ensure it is running, check the start of a log to indicate the same, and return a receipt with a valid job identifier. Given that we are connected to the same Flux broker, we can immediately verify if the job identifier is real, wait for the job to finish, and retrieve the final log from Flux programmatically for parsing. Additionally, the resource secretary software keeps track of all tool calls. With this combined information, we can inspect the state of the queue and tool calls to see if the agent submitted additional jobs or performed unexpected actions. A correct execution should have a request transformation as well as a submission, job information request, and log retrieval before returning a valid job identifier. Between each execution we completely clear the Flux queue.

\smallskip
\noindent{\bf Dispatch Prompt} 
\label{sec:dispatch-prompt}
The details of the dispatch prompt are important to ensure that the agent has clear instructions for its task. We instruct the agent to transform the user's textual intent into a cluster-specific submission command, preserving needs of the application. The agent is required to verify that the job is running for at least 10 seconds, which is achieved by looking at the log output to ensure no immediate errors have occurred. The agent must return a submission receipt, which includes a valid job identifier and reasoning. We can programatically verify the job existence, status and success, and retrieve the output log using the Flux Python SDK.

\smallskip
\noindent{\bf Evaluation} 
\label{sec:evaluation}
We will assess the degree to which the agent successfully completed the dispatch by way of the result of the execution and the agent behavior. To achieve this, we will look at the distribution of LAMMPS running times, and analyze the accuracy of the produced commands and job specifications. We can compare expected tool calls against actual, and the number of experiment runs that produced a job identifier with a completed run of LAMMPS. We will analyze the results to determine if there is an association between prompt style and job execution success.

\smallskip
\subsection{Multi-Cluster Dispatch Experiments} 
\label{sec:cloud-dispatch-experiments}

To demonstrate the ability of agents to use descriptive metadata to dispatch computational workloads we created a multi-cluster cloud experiment. We arrived at a design that would place agents (and required credentials) in user space, and allow for better performance and efficiency due to not needing to query an \gls{llm} \gls{api} within the scheduler. We create a separation of concerns between agentic components in user space and \gls{hpc} scheduling components in system space to ensure consistent and deterministic behavior of the latter. 

\smallskip
\noindent{\bf The Fluxq Project} 
\label{sec:fluxq}
We designed \emph{Fluxq}, a scheduling control plane with a queue manager that matches jobs to clusters using the Flux Framework graph-based scheduler, Fluxion. \emph{Fluxq} represents each cluster as a set of JSON Graph Format (JGF) graphs. A primary containment subsystem is used to describe traditional hardware such as nodes, sockets, cores, GPUs, and memory, and zero or more descriptive subsystem graphs can be registered alongside a cluster, allowing for matching on additional descriptive factors. We assume for our experiments and descriptive subsystems that clusters are homogeneous. Using this design, a user can make a work request with a quantitative \emph{resources} section in a job specification \cite{jobspec} and quantitative and qualitative needs (e.g., properties such as network fabric) in a requirements map from subsystem name to a resource section. Flux has support for an OR slot, a specification that allows the jobspec to find resources that match one request shape from a set of provided shapes. The OR slot allows for more flexibility in request syntax.

To support scalability and flexible subsystem addition and removal, each (cluster, subsystem) pair gets its own Fluxion instance in its own OS process. Each independent cluster can register with a name, and a manager type (e.g., Flux, Slurm, Kubernetes) that informs discovery of associated subsystems and configuration parameters. As an example, Kubernetes can accept a custom \emph{kubeconfig}, and will discover subsystems through programmatic access to clusters. Thus, a match for a single subsystem type can be run in parallel across clusters.

\begin{figure*}[h!]
    \centering
    \includegraphics[width=1.0\textwidth]{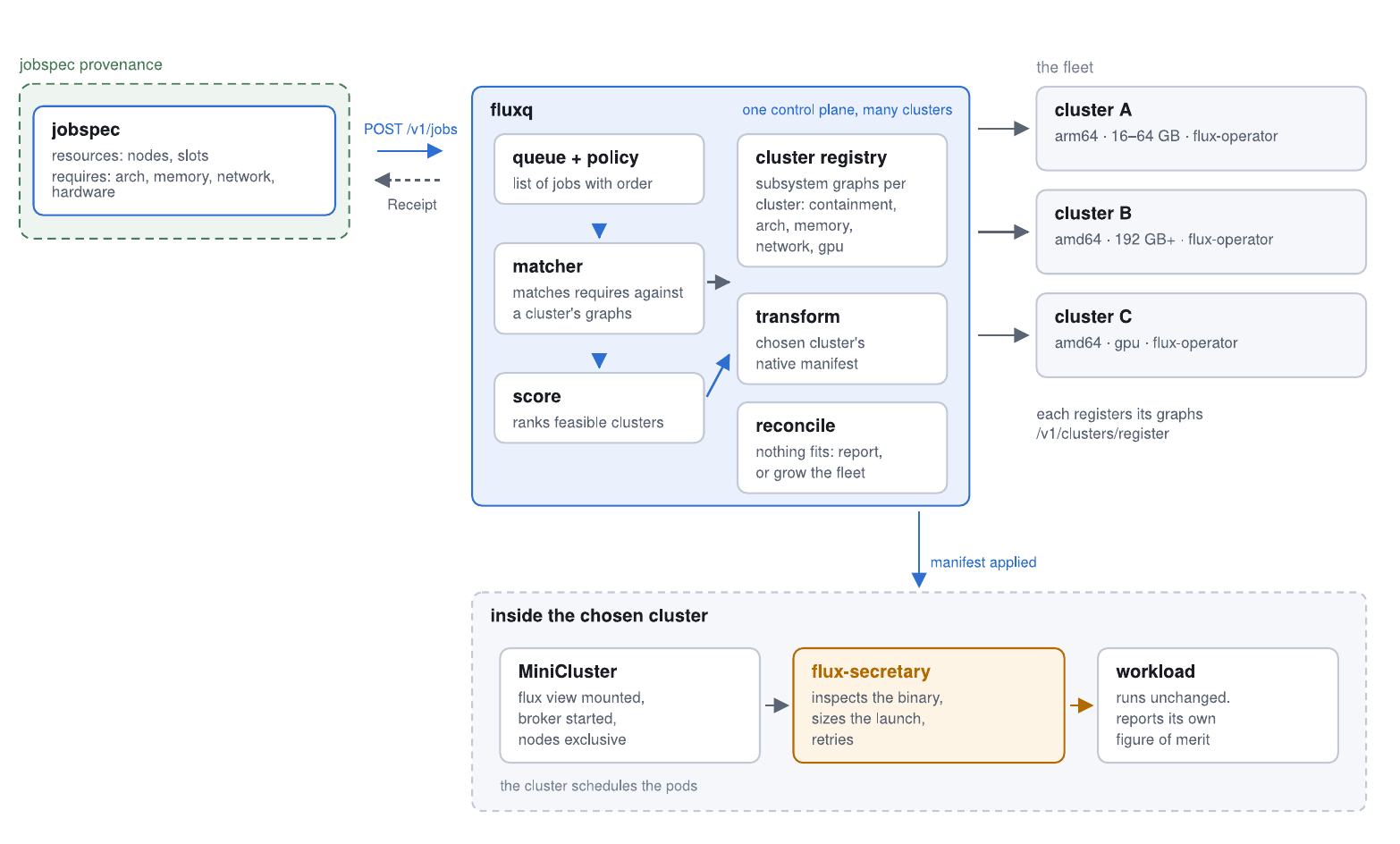}
    \caption{Fluxq Architecture. Clusters register resource subsystems to populate Fluxq graphs. A job specification is submit, matched against Fluxion, and scored. The job specification is transformed for the selected cluster and dispatched.}
    \label{fig:fluxq}
\end{figure*}

Interaction with \emph{Fluxq} is done programmatically via an authenticated RESTFul \gls{api}. Clusters can be registered, discovered, and receive submissions via programmatic endpoints. Execution is shown in Figure \ref{fig:fluxq} and proceeds as follows. The service is deployed and clusters register via a registration endpoint, declaring their backend type (e.g., the workload manager) and optionally with a request to discover subsystems. In a production setup, the cluster receives a secret to allow for further administrative interaction. A job specification is then submitted via a \emph{Fluxq} client library, matched against cluster subsystems in parallel using Fluxion, and scored. In the case of a tie the scores are randomly shuffled. The final scores are sorted and the top score is chosen as the selected cluster. The job specification is transformed for the selected cluster and dispatched.

\smallskip
\noindent{\bf Cloud Clusters} 
\label{sec:cloud-clusters}
To control for workload manager and performance variability due to node count, we aimed to deploy a uniform fleet of six clusters on two clouds, \gls{aws} and Google Cloud, that would have a consistent size of three nodes, and range in cost from 0.29 to 3.78 USD per hour.  Our fleet featured different networks, diverse hardware, and sizes (Table \ref{tab:fleet}). We attempted to use instance types with eight physical cores, a number that is common to instance types, allowing for instance type diversity with reasonable cost. The aim of the experiment would be to show agentic selection and dispatch with and without descriptive metadata represented as simple subsystems for architecture, network, and memory. The GPUs were not affordable based on usage time needed for the experiment. 

\begin{table*}[t]
\centering
\footnotesize
\setlength{\tabcolsep}{5pt}
\begin{tabular}{llcrlrrrllr}
\toprule
\textbf{Cluster} & \textbf{Cloud} & \textbf{Region} & \textbf{N} & \textbf{Instance} &
\textbf{vCPU} & \textbf{Cores} & \textbf{RAM} & \textbf{Arch} & \textbf{Network} & \textbf{\$/hr} \\
\midrule
gke-cpu            & GKE & us-central1 & 3 & e2-highcpu-16  & 16 &  8 &  16 & amd64 & gVNIC    & 0.40 \\
gke-mid            & GKE & us-central1 & 3 & e2-standard-16 & 16 &  8 &  64 & amd64 & gVNIC    & 0.54 \\
gke-arm            & GKE & us-central1 & 3 & t2a-standard-8 &  8 &  8 &  32 & arm64 & gVNIC    & 0.31 \\
gke-bigmem         & GKE & us-central1 & 3 & n2-highmem-32  & 32 & 16 & 256 & amd64 & gVNIC    & 3.78 \\
eks-arm-small      & EKS & us-east-1   & 3 & c7g.2xlarge    &  8 &  8 &  16 & arm64 & 15\,Gbps & 0.29 \\
eks-cpu-efa-bigmem & EKS & us-east-1   & 3 & m7i.4xlarge    & 16 &  8 &  64 & amd64 & EFA      & 0.81 \\
\bottomrule
\end{tabular}
\caption{Experimental fleet. \normalfont $N$ is nodes per cluster, RAM is nominal GiB per
node, and \$/hr is on-demand Linux compute for the whole cluster (US regions). \textbf{Cores} is physical cores.}
\label{tab:fleet}
\end{table*}

\smallskip
\noindent{\bf Artifact Secretary} 
\label{sec:artifact-secretary}
The \emph{artifact secretary} \cite{artifact-secretary} creates an isolated, scoped environment with an agent to explore a container and generate a manifest that describes applications inside. The discovery technique allows testing a large set of diverse containers with unknown contents. The project represents an improvement of our previous agentic work~\cite{fractale} that controlled each tool call to the \gls{llm} in a state machine. Instead, the artifact secretary uses an underlying library \cite{behalf} that provides an abstraction of a scoped agent provisioned by one of Google's Agent Development Kit (ADK), the Claude SDK, or Amazon's Bedrock AgentCore. The agent is given tools to list directories, read files, and run ELF inspection inside the container. Finally, the agent generates a JSON manifest that describes the application build.

\smallskip
\noindent{\bf Containers} 
\label{sec:containers}
We targeted set of 219 containers built for a usability study of cloud \cite{sochat2025usability}. We used the artifact secretary to discover executables and prepare manifests for submission to \emph{Fluxq} (Section \ref{sec:fluxq}). The selection resulted in 11 applications (Kripke, LAMMPS, AMG2023, HPL (LINPACK), miniFE, mixbench, OSU AllReduce (amd64 and arm64 builds), Quicksilver (amd64 and arm64), and STREAM). 
We will run the study under two conditions. In the first base condition, we derive a job specification that does not consider subsystem requirements. For example, a container manifest that warrants a \gls{hpc}-like fabric will not be known to the scheduler. In the second case, we will include subsystem requirements. We will run each paired comparison 10 times, for a total of 110 results per condition, and a total of 220 results. We hypothesize that application execution success and performance will improve when the scheduler has descriptive metadata available when choosing a placement. For our comparison, we will derive both pooled and paired differences as a representation of noise or latency added or removed by way of using descriptive metadata. Paired differences account for time in that the jobs between conditions are run back to back. Pooled allows for comparison even when one of a pair failed to run (e.g., an arm64 container being assigned to an amd64 cluster).

For our experiments, we use the AWS Bedrock backend with \emph{us.anthropic.claude-opus-5} as a model. The final manifest includes artifacts, each with an architecture, application name and binary, dependencies needed, and capability list. \emph{Fluxq} also exposes a vocabulary endpoint that dynamically derives subsystem options from the registered clusters. Using this vocabulary, associated tools can know what subsystems are registered, and options possible. Our vocabulary includes architecture (amd64, arm64), network (EFA, ethernet), and memory in ranges (0--16\,GB, 16--64\,GB, 64--192\,GB, 192\,GB+). We ask the agent to only specify a requirement for a container if the application warrants it. For example, the OSU benchmarks warrant an HPC-like fabric, and AMG is a memory bound application.

\smallskip
\noindent{\bf Jobspec Generation} 
\label{sec:information-granularity}
A manifest that describes the contents of a container next needs to be transformed into a job specification~\cite{jobspec} for the Fluxion graph scheduler. To match many clusters we use a strategy of least information: each stage is provided with the highest level of abstraction to complete the work without overfitting. As an example, a job specification that needs to match to a cluster only needs to define the number of nodes and subsystem needs for the application. The exact submission command depends on the eventual selected cluster. To generate the job specification, we instruct an agent to inspect a family of containers under a common repository (e.g., \texttt{ghcr.io/converged-computing/metric-lammps}), and discard variants that require resources that are not available in the cluster as determined by the subsystem vocabulary. The agent is instructed to select a container tag that meets these criteria, and use it to design a job specification that requires a particular count of exclusive nodes. % The job specification generation is written before running the full experiments to allow for the equivalent job specification to be submitted under the two conditions previously described. For the job specification, we instruct the agent to request exclusive nodes.

\smallskip
\noindent{\bf Job Lifecycle} 
\label{sec:queue-transform-dispatch}
A job specification that is submitted to \emph{Fluxq} will be queued, matched, scored, selected, transformed, and dispatched.  For the first step, submitted jobs enter a queue that is implemented using the River library in Go. Jobs are matched first to containment subsystems, and then descriptive. Descriptive requests are done as satisfy requests as they are modeled to represent consistent node features and do not need to reserve actual counts. For our experiment we also modeled the containment subsystem as a satisfy request to simplify the design. A satisfy request determines whether a resource graph can meet the needs of a job specification with property-based node features.  Although not used in our experiments, if an additional subsystem other than containment needed to be added with counts of a resource, \emph{fluxq} supports this use case.

Fluxion designates each cluster that can satisfy the containment and subsystem requests as a match.
Each match is scored by assigning points per matched subsystem, and the final scores are randomly shuffled and then sorted to break ties. The highest scoring cluster receives the allocation. In the absence of subsystems (the base condition) every cluster should be feasible and equally likely to be selected for a job. In the next stage, the job specification is transformed. A \emph{transform} interface implemented for the chosen cluster backend converts the job specification into a native artifact for the target manager. Two interfaces exist for transform: a deterministic template-based interface and an LLM-based one. We used the deterministic variant for our experiments to avoid additional latency introduced by an \gls{llm}. Dispatch is specific to the backend driver. For example, a Kubernetes manifest is applied via a specific kubeconfig context, while a Flux backend submits to a remote unique resource identifier. Each dispatch receives a submission confirmation that includes the clusters considered, rejected, and scored.

\smallskip
\noindent{\bf Execution} 
\label{sec:execution}
The execution step occurs when the resources are allocated. In our experiments, our clusters provisioned consistent Flux Framework MiniClusters in Kubernetes. We created software called the \emph{flux-secretary}~\cite{flux-secretary} that could receive application parameters, derive resources from the environment, and serve as an agent to successfully submit the work. In particular, the agent can submit and monitor the job, responding to errors up to a maximum number of retries (N=10). The execution step is closest to running the job, and thus the right level to finalize job submission details. The execution environment has immediate access to observable resources and specific workload manager launcher flags. The agent is allowed to change parameters for the launch (nodes, tasks, affinity, environment) and make parameter substitutions in the case of error. The agent is not allowed to change the executable or problem size, which would reflect a scientific objective. For each run, we capture the complete Flux Operator lead broker log that includes the cluster bootstrapping, the agent running and writing thinking to the output, and the application output or error.

We will record the matched cluster, the number of feasible clusters, which subsystems matched, the outcome, wall-clock duration, failure notes if applicable, and the full application log. We will derive performance \gls{fom} and outcomes from the log directly. We expect cluster selection to vary with and without descriptive metadata for the same job. The consistent use of Kubernetes with the Flux Operator and equivalent application containers is a means to reduce variation between different cluster selections.
\section{Results}
\label{sec:results}

We evaluated the performance and reliability of agentic dispatch across 432 combinatorial prompts. Our analysis focuses on the correctness of the generated job specifications, the performance of the dispatched LAMMPS workloads, and the agent behavior in the Kubernetes-based Flux environment.

\begin{figure}[t!]
    \centering
    \includegraphics[width=\columnwidth]{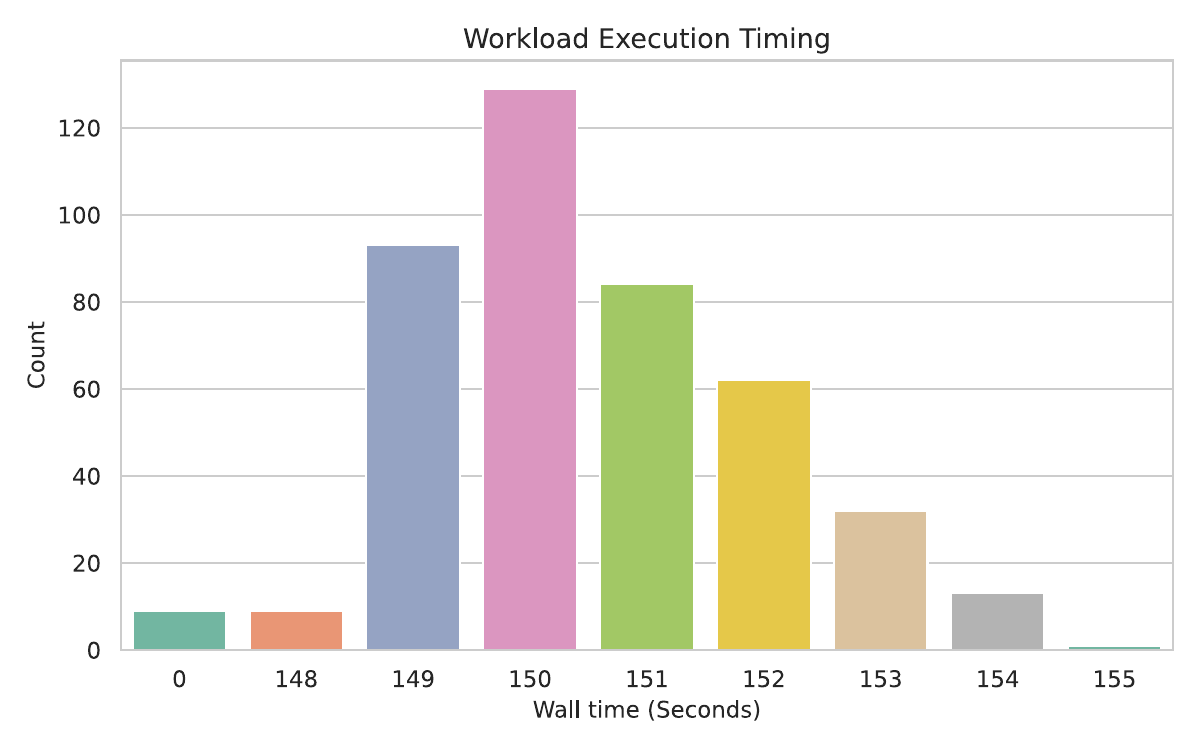}
    \caption{Distribution of wall-clock execution times for agentic versus manually executed LAMMPS. The agent's had a high success rate (97.9\%) to achieve a time comparable to a manual, gold standard.}
    \label{fig:execution-timing}
\end{figure}

\smallskip
\subsection{Dispatch Success Rate} 
\label{sec:dispatch-success-rate}
We compared agentic performance with manually executed gold standard runs (Figure \ref{fig:execution-timing}) and saw no significant difference in performance for LAMMPS wall times for the $32\times32\times16$ ReaxFF simulation. 97.9\% of executions (423/432) were successful to produce a comparable job time, and only 9 jobs failed to complete. Of the set that failed, one job had a \emph{False} placed in the command, and two jobs incorrectly concatenated the input flag with the target file (\emph{-in.reaxff.hns}). In simpler terms, the agent produced a period token instead of a space, which is akin to a typo. This result tells us that we should not only validate application problem sizes, but also application flags. For the remainder, the agent requested too few resources for the work, and it timed out. Specifically, the agent asked for 5 (N=1), or 64 (N=5) tasks all on one node. Observing the runs, we think the agent falls back to ``testing'' on one node given that an initial submission has issue, and erroneously reports its testing job identifier that eventually times out. More work is needed to work on validation of resource counts, as submitting jobs on different configurations of resources will require changing parameters and it cannot be a single hard coded set.

\smallskip
\subsection{Tool Calls} 
\label{sec:unexpected-calls}
\begin{figure}[hb!]
    \centering
    \includegraphics[width=\columnwidth]{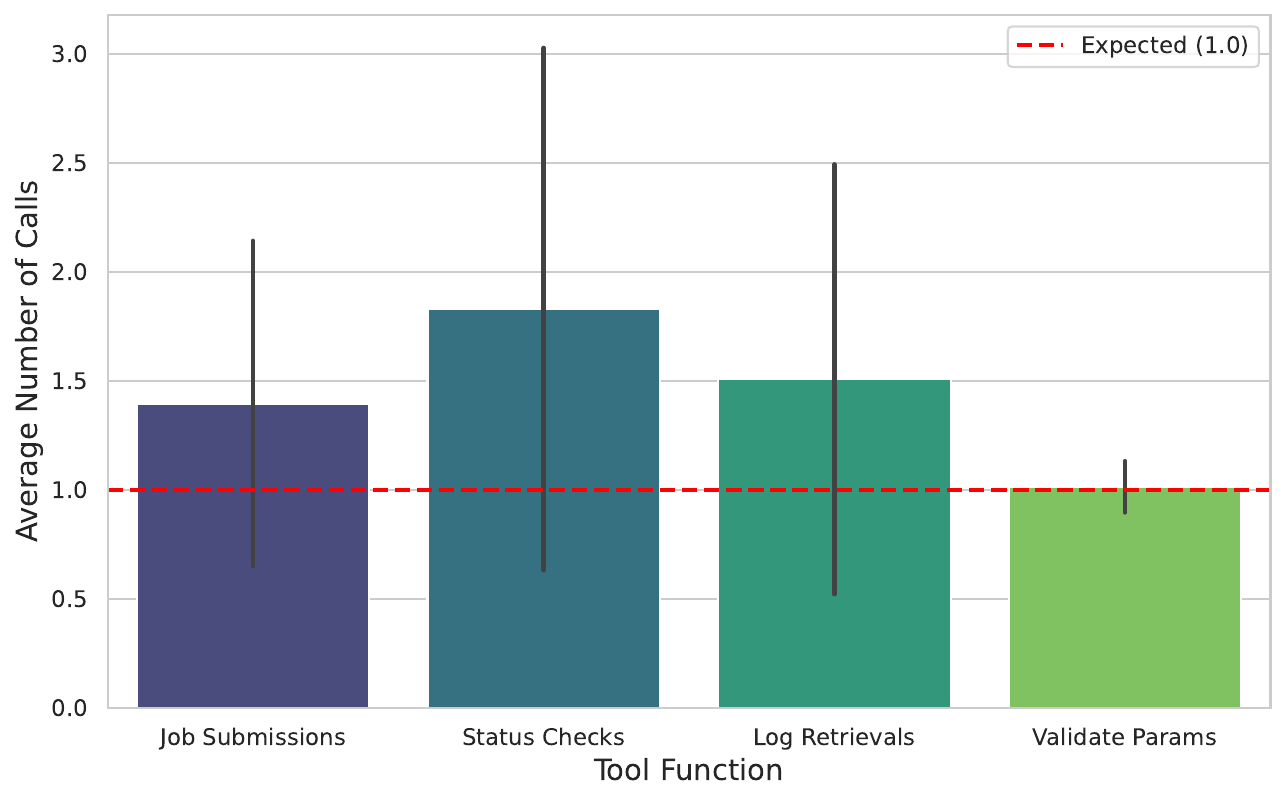}
    \caption{Average number of tool calls compared to the expected baseline of 1.0. The elevated submission reflects the agent's retry behavior.}
    \label{fig:tool-redundancy}
\end{figure}

A successful agentic execution for a short running application should have at least one submission, request for job info or status, request for logs, and LAMMPS parameter validation. Figure  \ref{fig:tool-redundancy} shows average number of tool calls. The agents reliably validated LAMMPS parameters once. A higher frequency of submission coincides with the agent retrying. It was common for an agent to request job info twice to verify job transition from \emph{PENDING} to \emph{RUN}. The higher rate of log requests also reflects the agents' decisions to inspect a log more than once. 

In Figure \ref{fig:tool-redundancy}, a small number of canceled jobs (N=2) can contribute marginally to these averages in that the agent would submit, monitor, and log more than once. Of the unexpected tool calls, while observing queue state (20.8\%) or resource status (12.0\%) was not required for a submission, it is not considered bad practice to explore the environment, and given a prompt that explicitly required it, the call would be needed. In that we did not instruct the agent to minimize calls, we do not consider this erroneous behavior. Observing the agent making the call requires more work to think about dispatch combined with fault tolerance in scheduling. While a job should not be dispatched to a cluster that is lacking in correct resources, the prompt and provided agent tools should support a means to notify a calling entity of an improper match and take appropriate action.

% \begin{figure}[h]
%     \centering
%     \includegraphics[width=\columnwidth]{images/unexpected_tool_calls.pdf}
%     \caption{Unexpected tool calls. With the exception of \emph{cancel\_job}, all unexpected calls are scoped under discovery and not considered erroneous.}
%     \label{fig:unexpected-calls}
% \end{figure}

\smallskip
\subsection{Environment and Workload Manager Flags} 
\label{sec:env-and-flags}
It was not common for the agent to set extra environment variables, with only 2.5\% (11/432) jobs having one or more variables. Of the 11, eight were for \emph{OMP\_NUM\_THREADS} to be set to 1. There were isolated cases of agents setting \emph{OMP\_PLACES} as \emph{threads}, \emph{OMP\_PROC\_BIND} as \emph{close,sockets} or \emph{true}, and \emph{OMP\_DISPLAY\_AFFINITY} as \emph{true}. While not required, there were 22 cases of the agent choosing to give the job a name. The agent was proficient at specifying the number of nodes, tasks, and duration for each of 99.1\%, 99.8\%, and 100\% of job commands. Less frequent but still common occurrences were to specify output or error, CPU affinity, and current working directory (N=103 and N=395, and N=374, respectively).

\smallskip
\subsection{Feature Analysis} 
\label{sec:feature-analysis}
We assessed the contribution of 5 features and four variants to job execution success. We generated an exhaustive prompt matrix to run LAMMPS (N=432) that resulted in 9 failures and 423 successful runs. The small number of failures deemed statistical testing not an option. Instead, we manually inspected the complete manifest of the run. For 7/9 cases, the error in the run was not related to the prompt, but instead was an issue with Flux losing contact with the shell or a job exception. Two of the nine cases were agent related. The agent incorrectly combined a LAMMPS flag with an argument, asking for \emph{-in.reaxff.hns} instead of \emph{-in in.reaxff.hns}.

\smallskip
\subsection{Cloud Dispatch Experiments}
\label{sec:cloud-dispatch-experiment-results}
Cloud dispatch experiments comprised 10 iterations of 11 applications under two conditions (with and without subsystems) for a total of 220 runs. Allocation placement differed between the base and subsystem conditions in 95 of 110 pairs.

\smallskip
\noindent{\bf Execution}
\label{sec:results-execution}
Stating requirements roughly doubled the proportion of jobs that executed, with 53/110 executing for the base condition and 96/110 with descriptive subsystems. Failure to execute was predominantly due to architecture mismatch for the base condition (43/57). Eight of the eleven applications for the subsystem condition had 100\% successful execution (10/10 successful completions). 

% \begin{figure}[t]
% \centering
% \includegraphics[width=\columnwidth]{images/execution.pdf}
% \caption{Execution. \normalfont Each application container (Y axis) and the number of total executions (X axis, full bar) versus those that were successful (X axis). Descriptive metadata provided by subsystems resulted in successful executions all sets of containers allocated. All failures are in the base condition.}
% \label{fig:execution}
% \end{figure}

The remainder of failures are not due to placement. Of the 14 incomplete subsystem runs, eight are Quicksilver, which faults inside \texttt{MPI\_Init} in libfabric on clusters without a matching fabric. It executed three times out of 20 across both conditions. The other five are LINPACK and mixbench, which fault in the same way. Closer inspection revealed that the agent likely gave up too early, not further exploring and making changes to the environment to enable the fabric. Techniques are needed to better inform the agent of the initial intent, and guarantee that more attempts are done. The agent was not aware of its maximum allowed retries.

\begin{figure}[t!]
\centering
\includegraphics[width=\columnwidth]{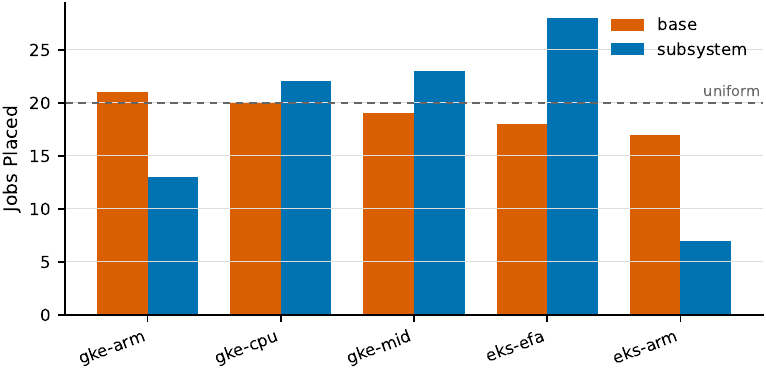}
\caption{Job placement by cluster. \normalfont There are 110 placements summed across clusters for each condition. The dashed line represents a uniform expectation.}
\label{fig:placement}
\end{figure}

\smallskip
\noindent{\bf Placement}
\label{sec:results-placement}
Without subsystems (the base condition) every cluster is feasible and is scored identically, so placement is decided by a random selection across clusters. We see this reflected in the uniform pattern of cluster selection, with each cluster receiving between 15 to 21 of 110 jobs (Figure \ref{fig:placement}). This uniformity establishes that the entire fleet was reachable and the matching and scoring worked as expected. In contrast, the subsystem condition is shaped by the applications' requests, ranging from placement of 7 to 28 per cluster out of 110. The \emph{eks-arm-small} cluster falls from 17 to 7 selections because only the two \texttt{arm64} applications may be placed there.

\begin{figure*}[t]
\centering
\includegraphics[width=\textwidth]{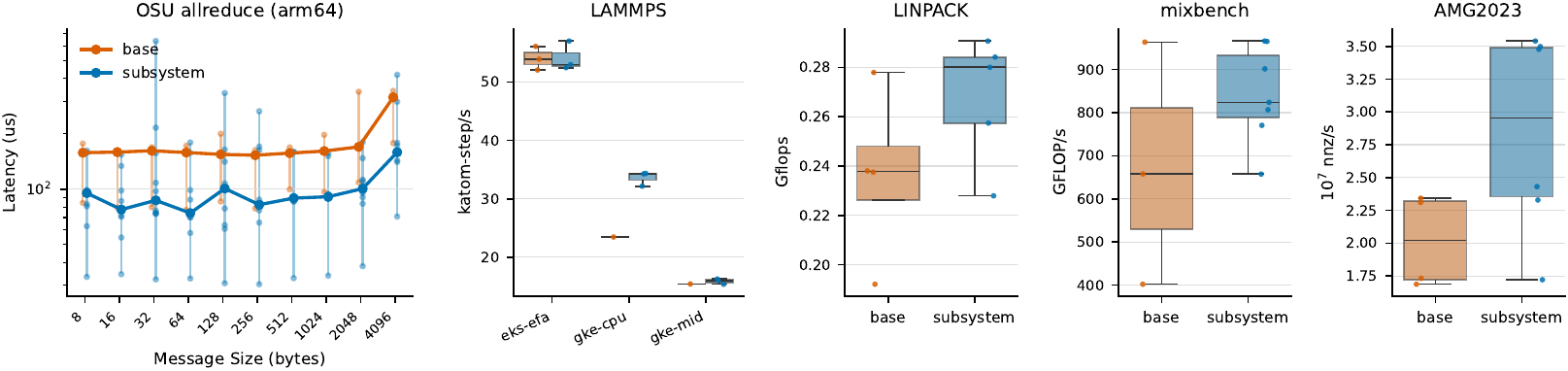}
\caption{Application performance examples. \normalfont MiniFE has a strong improvement in the subsystem condition. Improvements that are small reflect a relatively small and similar fleet.}
\label{fig:exemplars}
\end{figure*}

\smallskip
\noindent{\bf Performance}
\label{sec:results-performance}
Performance compares the subsystem condition against base control, not including the \emph{gke-bigmem} cluster due to a different number of cores. We classify results into three categories: subsystem advantage, base advantage, or no advantage. Examplars are shown in Figure \ref{fig:exemplars}. Applications that were faster or had higher \gls{fom} under the subsystem condition include MiniFE (3.32x), AMG2023 (1.46x), Stream (1.21x), LAMMPS, and Mixbench (1.11x). MiniFE is the most robust result both in magnitude and number of results to compare (6 base, 10 subsystem) as compared to Stream and Mixbench (3,9 and 2,5), respectively. Applications that had no measurable difference included Kripke, LINPACK, OSU AllReduce on amd64 (1.0, spread 2.6), and OSU AllReduce on arm64 (1.7, spread 2.4). Quicksilver completed every subsystem attempt but reported a \gls{fom} of zero on all of them. The problem size was too small for any segment to be tracked, meaning that the application launched successfully and did no measurable work. Thus, we could not include a variant of Quicksilver in our performance comparison as the base condition, and can only report on 10 measurable applications.

For application performance we observe three patterns, and provide exemplars (Figure \ref{fig:exemplars}).  The first pattern is a cluster with a clear performance improvement. MiniFE ran 3.3x faster with subsystem selection as compared to the base condition. The second pattern is marginal difference that is likely attribed to noise (e.g., Linpack and Kripke). For the \code{arm64} OSU benchmark the subsystem condition has lower latency at all ten message sizes, by a median factor of 1.66 at the smallest size. 
  
\smallskip
\noindent{\bf Agent availability}
\label{sec:results-agent}
The model \gls{api} became unavailable during 5 of the 220 runs. All five fell back to execute the original deterministic command and none of them completed. Exponential backoff should be used in future experiments.

\section{Discussion} 
\label{sec:discussion}

We have performed experiments that present evidence that using \gls{llm}s with agentic interfaces leads to successful job execution. We demonstrate through end-to-end experiments how level of detail in a job specification corresponds to levels of orchestration. An initial submission that needs to match many clusters best defines the high-level resources and scale. Once received by a cluster, an agent can successfully generate a fine-grained command. Over-specificity too early will prune away potential candidates, and under-specificity later may result in lower performance. The higher level of detail is a representation of intent, while the lower-level detail specifies the mechanism to achieve it. Portability is achieved by transporting job specifications around the scheduler to best describe intent over mechanism. Our results provide early evidence that descriptive and agentic orchestration presents promise for executing scientific workloads. % It is possible to describe what a workload needs, match it to a fleet, and decide how to run it more dynamically once it is allocated resources.

\smallskip
\subsection{Environment and Workload Manager Flags} 
\label{sec:env-and-flags-discussion}
We learned that it was important to provide granularity about binary paths, and to omit flags such as specific queue or bank. We observed this behavior with building containers in our previous work \cite{sochat2025agentic}, where not documenting or providing an entrypoint to execute an application command was a source of error. Hallucination can result from both using lower-capability models and not providing sufficient information to the agent. Validation of the application-specific commands is important, along with ensuring that workload manager and application flags are correctly placed. In our testing, the agent would often interleave flags, leading to erroneous execution due to invalid syntax.

\smallskip
\subsection{Transform Limitations} 
\label{sec:transform-limitations}
Despite a 97.9\% success rate, the failures we observed highlight several limitations with respect to our current tool choices, and present opportunities for improvement. First, the agent incorrectly concatenating the input flag to the file name suggests that additional tools are needed to validate the application command. An agent should ask any binary for \emph{--help} to sanity check how it has formulated the arguments. We will consider an adversarial approach of one agent checking and providing feedback to a main execution agent. A second limitation in our work is the choice of models. We used Gemini due to token availability. AI agent non-determinism means that using different models, or versions of the same model, could lead to different results. Finally, we recognize the challenge of implementing application- or workload manager specific tools. A future software engineer role may be as a discovery tool designer for function endpoints intended for use by agents. Other limitations are assuming a pre-authorized environment, and reliance on a synchronous polling model. We have implemented event subscriptions for our \emph{mcpserver} and plan to use and test them in future work.

% Removing this because the flux-secretary essentially does this.
% Additionally, we identified a limitation in our agents' recovery behavior. After a failure, an agent would want to fall back to testing behavior on a smaller number of nodes, and then report the smaller, incorrect result as the submitted job. These jobs would eventually timeout because the problem size was too large to be completed during the specified walltime. To resolve this and other issues that might arise, an agent needs to be empowered to debug and better deal with uncertainty. Execution could be returned to a calling agent or debugging agent to get feedback on the issue before trying again, an approach we have shown has relative success in previous work \cite{sochat2025agentic}.

% The agent is required to check the job status and log output manually to verify success, which increases the total deliberation time and the number of required tool calls. A better approach is event driven, where an agent can handle several tasks at once and receive job notifications as they come. We have implemented event subscriptions for our \emph{mcpserver} and plan to use and test them in future work. For authorization, work is underway with the Genesis Mission to work on this problem for multi-tenant systems across centers \cite{Gil2025Genesis}.

\smallskip
\subsection{Descriptive metadata to inform multi-cluster dispatch}
Requirements did not reliably make jobs faster. Rather, descriptive metadata helped by excluding machines that could not run the job at all. The subsystem selection prevented the worse case outcome and was not a consistent performance optimization. Our subsystem space was overly simple, reflecting basic networks, architecture, and memory for a small set of clusters that had little difference across dimensions. Further work is needed to improve modeling of subsystem variability, and include both GPUs and application definitions with clear preferences for particular features. We also tried to write job specifications to maximize generalizability, often requesting \texttt{ethernet} or \texttt{efa} for the network. In practice, this meant that applications that would perform better with a fabric were placed on a Google Cloud cluster without one. There is a tradeoff between portability and performance \cite{sochat2025usability}. A more portable application can run on more clusters, but often at the cost of performance.

\smallskip
\subsection{Cloud Dispatch Limitations} 
\label{sec:cloud-dispatch-limitations}
A limitation of our multi-cluster dispatch work is that intent did not reach the agent. The \emph{flux-secretary} receives an allocation and a command, and although it is given tools to inspect the binary it will launch, it is not aware of the larger picture that formed the execution. In future work, we will consider better passing forward this intent. Our experiments can also be extended by testing different algorithms for scoring, accounting for queue depth and cost, and adding variability to node counts and problem sizes.  The setup might support a job failure that is returned to the scheduler to be rematched.  % Feasible clusters are ranked by counting matched subsystems and do not account for queue depth or cost. The agent should have been configured to retry with exponential backoff and jitter on \gls{api} failures.  Our placement and execution results retain all six clusters, 220 runs and 110 pairs. Despite a small number of clusters and applications, the entire orchestration took one of our authors over 48 hours of full serial experiment running, observing, and testing time, a task that would have been not cost effective using larger sizes or expensive resources like GPUs.

\smallskip
\subsection{Future Work} 
\label{sec:future-work}
Resource unavailability due to contention demands that workloads are run in suboptimal environments. Such different configurations will require not just translation and dispatch between managers, but also translation of an initial job intent to a different set of resources or compiled software. A simple example is running a workload that was intended for CPU on GPU machines. This task that focuses on goal-oriented translation is different than a transformation of static resource requests, and is a much more challenging task. We expect agents to need access to models that provide information about historical performance data, and other policy or optimization metrics of interest.

Our initial multi-cluster cloud work lays the foundation for robust and flexible agentic workload orchestration. We intend to expand on number and types of tested environments, and better define an application and resource vocabulary for subsystems.

\section{Conclusion}
Agentic frameworks are central to the future of the \gls{hpc} community and the larger software ecosystem. Autonomous, converged \gls{hpc} infrastructure that can intelligently negotiate job workloads is becoming feasible.  From autonomous task execution to failure recovery and AI-supported scheduling techniques, these powerful transitions have and will continue to change the day to day work of computational scientists and engineers. We anticipate a new era of faster, more accurate, and more accessible computing for the next generation of \gls{hpc}.

\section*{Acknowledgment}
Thank you to Livermore Computing for supporting us in all respects. This work was performed under the auspices of the U.S. Department of Energy by
Lawrence Livermore National Laboratory under Contract DE-AC52-07NA27344 and
was supported by the LLNL-LDRD Program under Projects No. 24-SI-005 (LLNL-CONF-2023057).

\bibliographystyle{IEEEtran}
\bibliography{references}

% \newpage
% \include{sections/artifacts}

\end{document}